\documentclass{article}
\usepackage{spconf,amsmath,graphicx}
\usepackage[misc]{ifsym}
\usepackage{amsfonts}
\usepackage{tabularx}

\title{Do self-supervised speech and language models extract similar representations as human brain?}
\name{Peili Chen$^{1}$, Linyang He$^{2}$, Li Fu$^{3}$, Lu Fan$^{3}$, Edward F. Chang$^{4}$, Yuanning Li$^{1 \text{ \Letter}}$
\thanks{This work is supported by NSFC General Program 32371154 and Shanghai Pujiang Program 22PJ1410500. \Letter \ Correspondece to: Yuanning Li (liyn2@shanghaitech.edu.cn).}
}
\address{$^{1}$School of Biomedical Engineering, ShanghaiTech University, Shanghai, China.\\
$^{2}$Department of Electrical Engineering and Computer Science, University of Michigan, Ann Arbor, MI, USA.\\
$^{3}$JD AI Research, Beijing, China.\\
$^{4}$Department of Neurological Surgery, University of California, San Francisco, CA, USA.
}
\begin{document}
%
\maketitle
\begin{abstract}
Speech and language models trained through self-supervised learning (SSL) demonstrate strong alignment with brain activity during speech and language perception. However, given their distinct training modalities, it remains unclear whether they correlate with the same neural aspects. We directly address this question by evaluating the brain prediction performance of two representative SSL models, Wav2Vec2.0 and GPT-2, designed for speech and language tasks. Our findings reveal that both models accurately predict speech responses in the auditory cortex, with a significant correlation between their brain predictions. Notably, shared speech contextual information between Wav2Vec2.0 and GPT-2 accounts for the majority of explained variance in brain activity, surpassing static semantic and lower-level acoustic-phonetic information. These results underscore the convergence of speech contextual representations in SSL models and their alignment with the neural network underlying speech perception, offering valuable insights into both SSL models and the neural basis of speech and language processing.

\end{abstract}
\section{Introduction}
\label{sec:intro}
Self-supervised learning (SSL) models have achieved state-of-the-art performance in a variety of speech and language tasks. SSL models are extremely effective in extracting the critical content and structural information from massive text or speech datasets, such as phonetics, syntax, semantics, and rich contextual information \cite{hsu2021hubert, peters-etal-2018-deep, pasad2021layer}, using generative \cite{radford2019language}, contrastive \cite{baevski2020wav2vec} and masked-prediction \cite{hsu2021hubert, kenton2019bert} objectives. However, the extent to which the core representations and computations in these SSL models align with the processes of speech and language in the human brain remains unclear.


Researchers have investigated the correlation between SSL models and brain activity, typically learning the brain-SSL alignment \cite{qiu2023can, wang2022open}, or involving linear mapping from SSL models' internal embeddings to the corresponding neural responses of speech and language, and assessing the alignment performance based on the neural activity prediction accuracy \cite{schrimpf2020integrative}. In \cite{millet2022toward} and \cite{li2022dissecting}, embeddings of transformer-based self-supervised speech models more accurately predict neural response to speech in auditory cortex, compared to traditional feature encoding models using heuristic sets of acoustic and phonetic features \cite{mesgarani2012selective}. In \cite{goldstein2022shared, schrimpf2021neural}, transformer models trained on the next-word prediction task predict the neural responses to speech and language with high accuracy across different datasets and imaging modalities. While these studies show a high brain alignment between SSL models and the brain, they usually model neural activity using only speech or language models. It remains unclear if the speech model and language model correlated to the same aspects of neural processing or to different levels of representations located in the same cortical areas, especially considering these models are trained on very different data modalities and tasks. 

In this work, we directly investigate the shared representations between speech and language SSL models and the corresponding brain activity by evaluating the brain encoding performance of speech models and language models.




Our main contributions are shown as follows: 

1. We demonstrate that Wav2Vec2.0 \cite{baevski2020wav2vec}, a representative SSL speech model, and GPT-2 \cite{radford2019language}, a representative SSL language model, both accurately predict speech response in the human brain, and their brain predictions are significantly correlated. 

2. Our results further highlight feature space similarities between these SSL models, with the shared representation explaining most auditory cortex neural activity variance.

3. Shared information contributing to brain encoding comprises primarily contextual, followed by static semantics and acoustic-phonetic, suggesting transformer-based SSL models learn linguistically relevant information similar to the brain.

\begin{figure}[htb]

\begin{minipage}[b]{1.0\linewidth}
  \centering
  \centerline{\includegraphics[width=8.5cm]{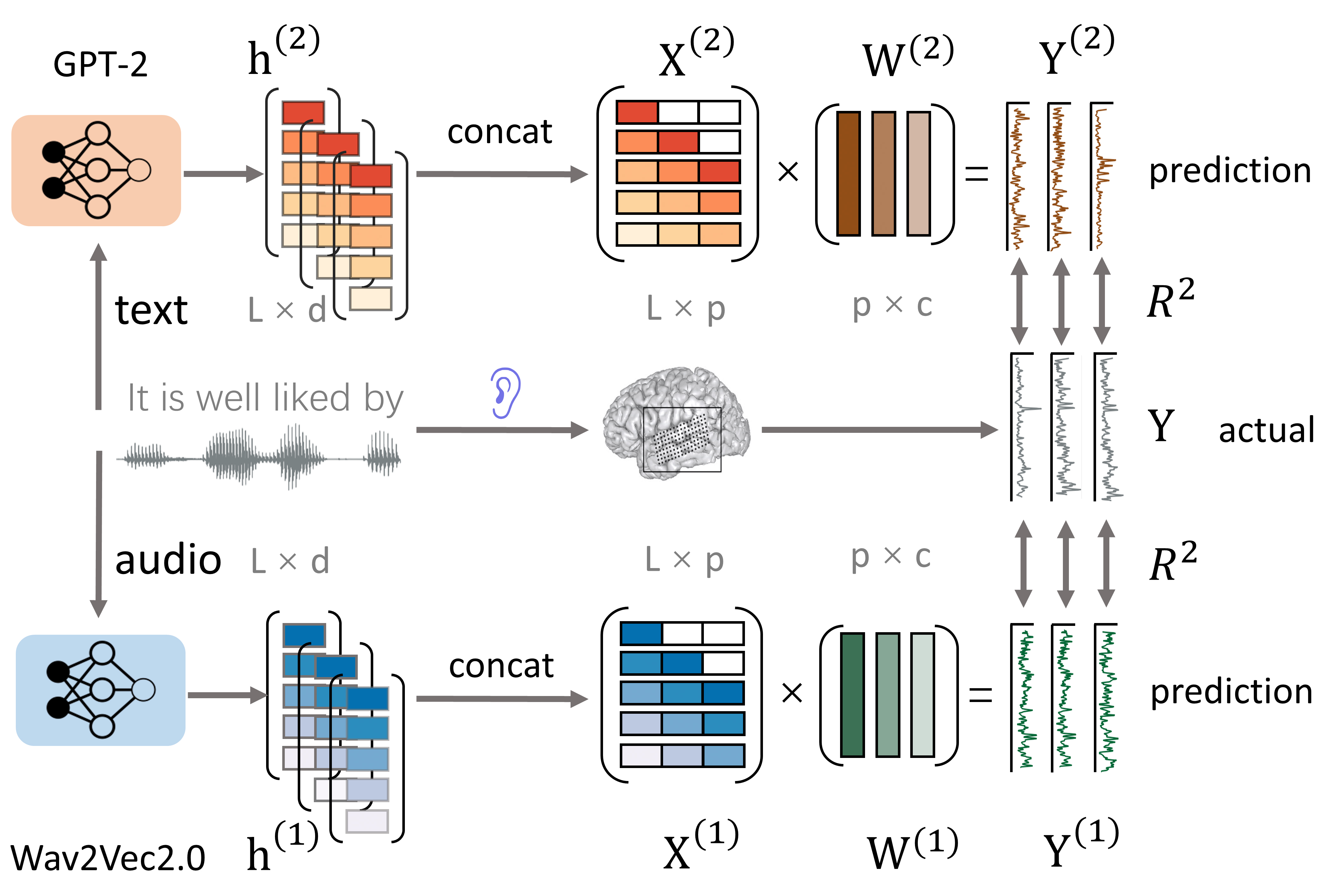}}
\end{minipage}
\caption{Predicting brain activity using SSL models.}
\label{fig1}
\end{figure}

\section{ANALYSIS METHODS}
\label{sec:format}


%
%
\subsection{Overview of the analysis}
We employed the temporal receptive field (TRF) model \cite{theunissen2001estimating}, a widely utilized brain encoding model in existing literature \cite{millet2022toward, li2022dissecting, goldstein2022shared, schrimpf2021neural}. In Figure \ref{fig1}, we assessed the neural prediction performance of embeddings from both Wav2Vec2.0 and GPT-2. Additionally, we derived shared components through canonical correlation analysis (CCA) from these model embeddings. We then evaluated the content of the shared information between SSL models using these shared embeddings in the same neural encoding pipeline.

\subsection{Deep neural encoding model using SSL speech and language models}
First, we extracted the SSL model activation of the TIMIT data. Audio waveforms and corresponding text transcriptions served as input for the speech and language models, respectively, yielding activation sequences $\mathbf{h} = [\mathbf{h}(1),...,\mathbf{h}(L)]^\text{T} \in \mathbb{R}^{L\times d}$, where $d$ represented the embedding dimension (768 for Wav2Vec2.0; 1600 for GPT-2 XL) and $L = 58943$ referred to the sequence length encompassing the entire audio/text. For GPT-2, we ensured alignment by repeating embeddings within the duration of single words to match the length of Wav2Vec2.0's embedding sequences. To ensure temporal correspondence, we resampled the ECoG signal to match the length of the SSL embedding sequences, with $y_i(t)$ representing the ECoG recording from the $i$-th electrode at time instances $t=1,...,L$, and $i=1,...,c$, where $c$ denoted the total number of speech-responsive electrodes.


Then we fit the L2-regularized linear regression model for each individual electrode
\begin{equation}
    \underset{\mathbf{w}_i}{\text{argmin}} ~~ \frac{1}{L}\sum_{t=1}^L\left[\sum_{\tau=0}^{w-1}\mathbf{w}_i(\tau)^\text{T}\mathbf{h}(t-\tau) + b_i - y_i(t)\right]^2 + \lambda_i \|\mathbf{w}_i\|_F^2
\end{equation}

Here the regression weight matrix $\mathbf{w}_i = [\mathbf{w}_i(0)^\text{T},...,\mathbf{w}_i(w-1)^\text{T}] \in \mathbb{R}^{d \times w}$ was the "temporal receptive field" with window length $w$ fixed at 400 ms, $\lambda_i$ was a tunable weight of L2 regularization determined through cross-validation. This formulation could be further simplified in matrix form
\begin{equation}
    \underset{\mathbf{W^{(s)}}}{\text{argmin}} ~~\frac{1}{L}\|\mathbf{X^{(s)}W^{(s)}} + \mathbf{b^{(s)}} - \mathbf{Y}\|_F^2 + \|\mathbf{W^{s}}\Lambda^{1/2}\|_F^2
\end{equation}
where $\mathbf{Y}\in \mathbb{R}^{L \times c}$ was the ECoG data matrix, $\mathbf{X^{(s)}} \in \mathbb{R}^{L \times p}$ was the concatenated embedding matrix with delayed window ($p = w \cdot d$), $\mathbf{W^{(s)}} \in \mathbb{R}^{p \times c}$ was the concatenated weight matrix, $\mathbf{b^{(s)}} = [b_1,...,b_c]^{\text{T}}$ was the intercepts of the ridge regression models, and $\Lambda = \text{diag}\{\lambda_1,...,\lambda_c\}$, $s=1,2$ corresponded to Wav2Vec2.0 and GPT-2 respectively. 

Such models allowed us to predict the neural activity from the embedding features of the SSL models in a window of time preceding the neural activity.



\subsection{Brain-prediction performance metric}
We quantified the neural encoding model performance by computing the regression $R^2$ between predicted neural response $\widehat{\mathbf{y}_i}(t) =\sum_{\tau=0}^{w-1}\mathbf{w}_i(\tau)^\text{T}\mathbf{h}(t-\tau) + b_i$ and the actual neural response $\mathbf{y_i}(t)$, defining it as the "prediction score." For each ECoG electrode, the prediction score of a certain SSL model was the optimal $R^2$ maximized over individual transformer layers. We estimated the noise ceiling for each electrode using repeated trials of the same sentences to establish the upper limit of explainable variance. The normalized prediction score was obtained by dividing the prediction score by the noise ceiling. 

\subsection{Analyze shared feature representations using Canonical Correlation Analysis}
\label{ref:3.2}
Canonical Correlation Analysis (CCA) is a multivariate statistical technique to explore the correlation between two groups of variables \cite{hotelling1936cca}. 
To evaluate how much explainable variance is contributed by the information shared by Wav2Vec2.0 and GPT-2, we employed CCA to encode this shared information into canonical variables. The top 50 pairs of canonical variables (CCs) captured the most common information shared between Wav2Vec2.0 and GPT-2, while the remaining CCs represented the residual information unique to each model. These canonical variables served as input for the deep neural encoding model to predict neural activity.

Linguistic theory posits three key dimensions within speech-language: acoustic-phonetic information, static semantics, and structural/contextual information \cite{brennan2022language}. We further modeled these three aspects as three models and used them to dissect the shared representations between SSLs. Inspired by \cite{pasad2021layer}, we applied CCA to analyze the static semantic information and acoustic-phonetic information encoded in the shared representation between SSLs, probing with GloVe and mel spectrum respectively. For static semantics, we employed GloVe embeddings \cite{pennington2014glove}, tailored to capture semantic meaning independently of context. To represent acoustic-phonetic features, we utilized the mel spectrum. Contextual information was captured through residual context embeddings, as introduced by Toneva et al. in \cite{toneva2022combining}. This approach extracted contextual information from GPT-2 by regressing out first-layer embedding features from the embeddings of the intermediate layer, resulting in contextual representations excluding lexical-level features from the first layer.


\begin{figure}[t]

\begin{minipage}[b]{1.0\linewidth}
  \centering
  \centerline{\includegraphics[width=7.5cm]{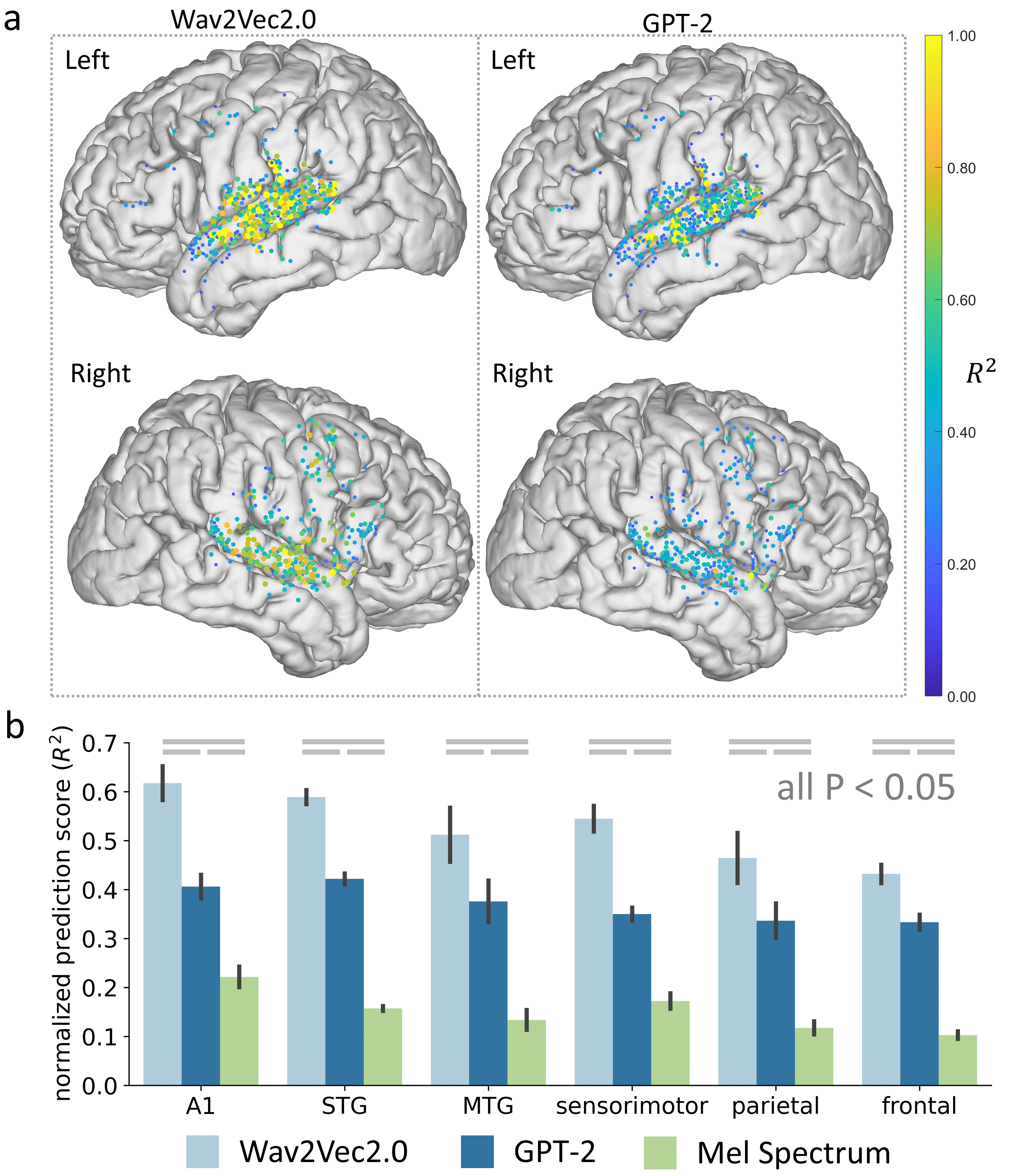}}
\end{minipage}

\caption{\textbf{a.} Brain prediction scores of Wav2Vec2.0 and GPT-2 on the left and right brain hemispheres, each dot is a single ECoG electrode. \textbf{b.} Average normalized prediction score over the electrodes in brain areas related to speech perception, compared to a baseline prediction model using mel spectrum features. (P values computed from paired t-test, Bonferroni correction)}
\label{fig2}
\end{figure}

\begin{figure}[thb]
\begin{minipage}[b]{1.0\linewidth}
  \centering
  \centerline{\includegraphics[width=7.5cm]{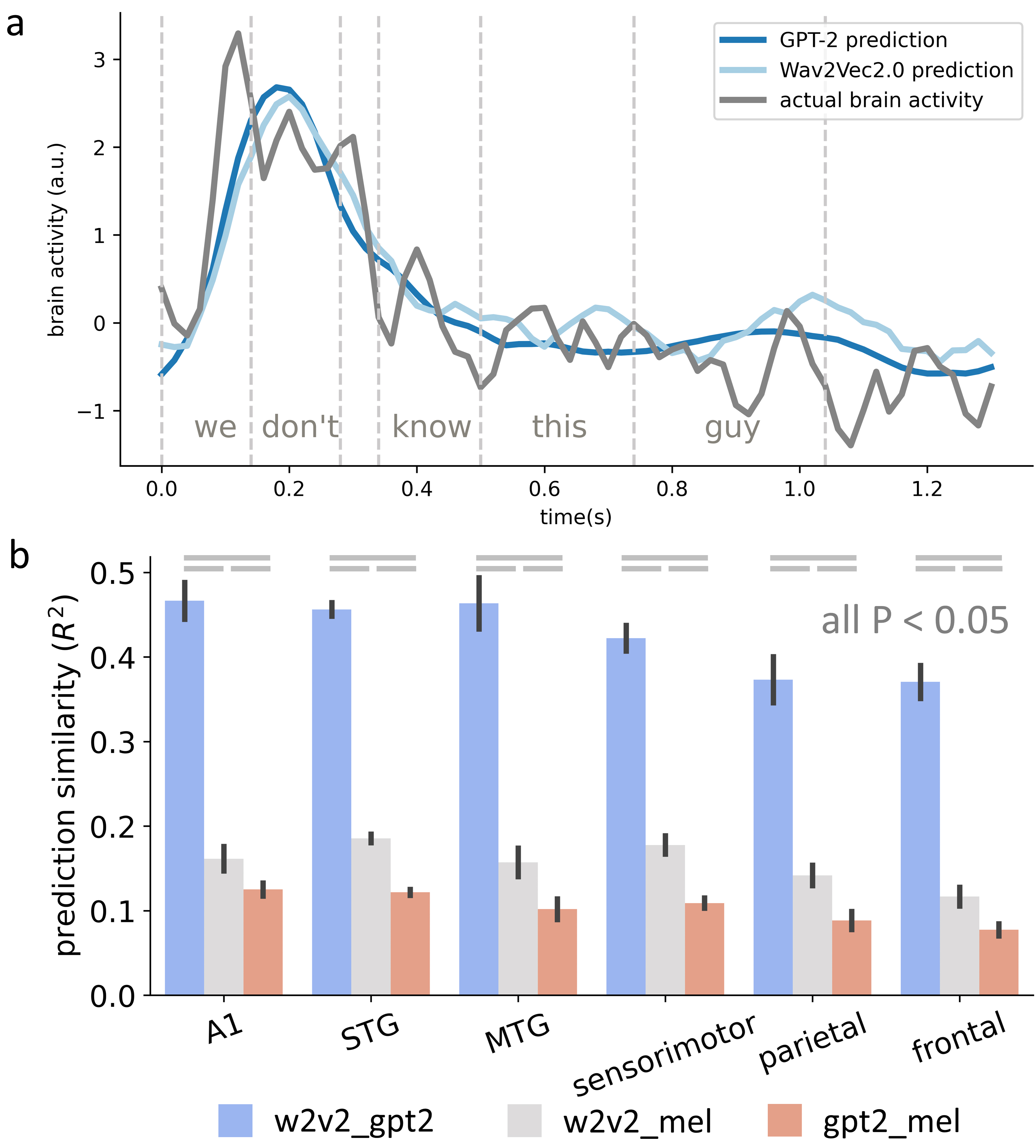}}
\end{minipage}
\caption{\textbf{a.} An example of the actual and SSL-predicted neural activity corresponding to a sentence from one ECoG electrode in STG.  \textbf{b.} The similarity between the brain activity predictions of Wav2Vec2.0 (w2v2), GPT-2 (gpt2) and mel spectrum (mel), quantified by shared variance ($R^2$). (P values computed from paired t-test, Bonferroni correction)}
\label{fig3}
\end{figure}

\section{Experiment results}
\label{sec:pagestyle}\subsection{Experiment setup}
{\bf Neural recording data.} We conducted ECoG recordings on 16 native English-speaking subjects while they listened to 599 sentences from the TIMIT corpus \cite{garofolo1993darpa}.\footnote{The experiment was approved by the Institutional Review Board at UCSF. Participants gave full consent before the experiments.} ECoG offers high spatiotemporal resolution and signal-to-noise ratio compared to non-invasive methods like fMRI or EEG. Electrode grids were implanted during epilepsy treatment, covering the cortical language network in temporal, parietal, and frontal lobes. We analyzed 822 speech-responsive electrodes out of a total of 4388, primarily located in the superior temporal cortex (STG), sensorimotor cortex (SMC), and supramarginal gyrus (SMG) (see Fig. 2a).

\noindent{\bf Models.} We employed two pre-trained models: Wav2Vec2.0 \cite{baevski2020wav2vec}, a self-supervised speech model, and GPT-2 \cite{radford2019language}, a language model. Wav2Vec2.0 utilizes a 12-layer Transformer encoder for self-supervised training on 960h Librispeech data \cite{librispeech}, generating audio sequence embeddings from 16 kHz raw speech waveforms. GPT-2, on the other hand, features 48 Transformer layers and was trained on a vast text corpus before fine-tuning for versatility. These models were directly used without further neural data training, rendering them agnostic to neural speech representations.

\begin{figure}[thb]
\begin{minipage}[b]{1.0\linewidth}
  \centering
  \centerline{\includegraphics[width=7.5cm]{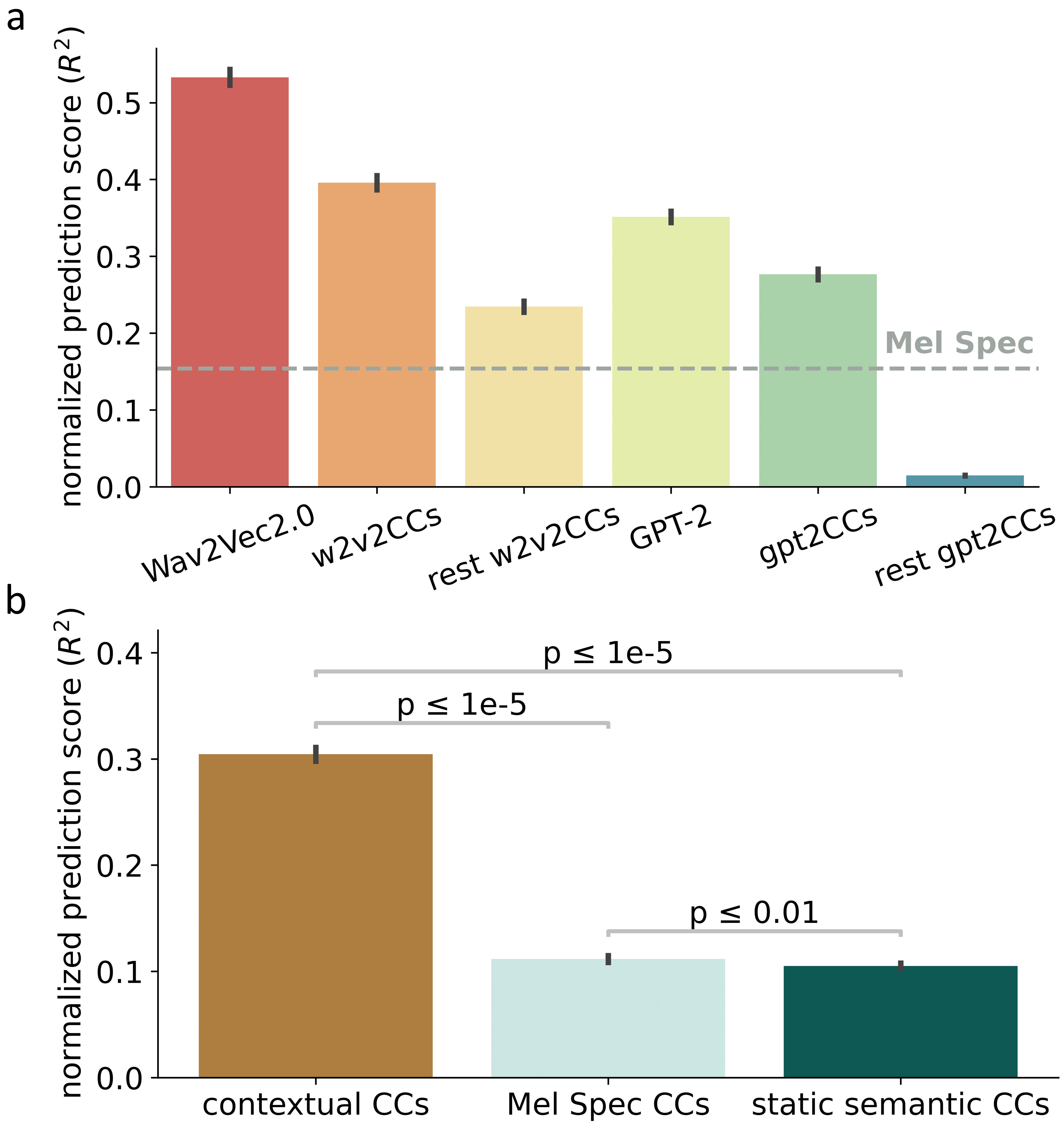}}
\end{minipage}
\label{fig4}
\caption{\textbf{a.} Normalized prediction score of the top 50 canonical variables (*CCs) and the remaining canonical variables (rest *CCs) between Wav2Vec2.0 (*=w2v2) and GPT-2 (*=gpt2), compared to their original prediction scores (Wav2Vec2.0 and GPT-2), and mel-spectrum baseline (dashed line). \textbf{b.} The brain prediction performance computed from the decomposition of the shared CCs (gpt2CCs). The decomposition was computed from CCA between gpt2CCs and three models: GloVe (static semantic CCs), mel spectrum (Mel Spec CCs) and residual context (Contextual CCs). (P values computed from paired t-test, Bonferroni correction)}
\end{figure}

\subsection{Analysis results}
{\bf Speech and language SSL models both predict brain response with high accuracy.} We first mapped the two SSL model embeddings to brain activity and assessed the performance using the brain prediction score.

Fig. 2 illustrates the brain prediction scores for Wav2Vec2.0 and GPT-2 models across individual cortical electrodes. Both SSL models significantly outperformed the baseline linear Spectro-temporal receptive fields (STRF) model with mel-spectrogram features, a widely used neural encoding model in the literature \cite{mesgarani2012selective, theunissen2001estimating} (Fig. 2b). For both models, high-scoring electrodes were primarily located in the bilateral primary auditory cortex (A1) and superior temporal gyrus (STG), areas associated with auditory processing during speech and language perception \cite{hickok2007cortical}. This is also consistent with the theory that STG plays a central role in extracting meaningful linguistic features from speech \cite{yi2019encoding}. 


These results validate the findings in the literature. Furthermore, Wav2Vec2.0 was consistently higher than GPT-2 across the cortex, suggesting that speech SSL model better fits the speech neural responses.

\parskip 0.1in
\noindent{\bf The speech model and the language model correlate to the same aspect of neural response to speech.} 
We then tested whether the speech model and the language model explain the same aspect of neural response to speech. In Fig. 3a, predictions from both SSL models displayed similar dynamics and tracked the actual brain activity for a sample sentence. Fig. 3b quantified their prediction similarity, revealing that the two models shared nearly 50\% of the prediction variance in the auditory cortex (A1, STG, \& middle temporal gyrus(MTG)). This similarity between the SSL models consistently exceeded their shared similarity with the mel-spectrogram predictions. 

\parskip 0in
On the SSL embeddings side, we applied Canonical Correlation Analysis (CCA) to extract shared information between Wav2Vec2.0 and GPT-2 embeddings. Using the first 50 Canonical Components (CCs) for neural activity prediction, we found that this subset effectively captured the shared brain prediction between SSL models (Fig. 4a).

The performance of the residual CCs for Wav2Vec2.0 surpassed that of GPT-2, indicating that Wav2Vec2.0 contributes unique information in addition to the shared aspects. GPT-2's unique information contributed little to explainable variance in the brain (Fig. 4a). This also explained why Wav2Vec2.0 outperformed GPT-2 in brain prediction (Fig. 2b).

Overall, these results imply significant information sharing between the brain, especially STG, and both Wav2Vec2.0 and GPT-2. We confirmed that these SSL models correlated to the same aspect of neural response to speech.

{\parskip 0.1in
\noindent{\bf Contextual information of speech accounts for the shared information between speech models, language models, and the brain.} 
Finally, we examined the exact content of the shared representations between the speech and language models. We considered three aspects of speech representation \cite{brennan2022language}: form (acoustic-phonetic information), meaning (semantics), and structural/contextual information. To represent these aspects, we utilized mel spectrum, GloVe embedings \cite{pennington2014glove}, and residual context embeddings respectively.}

\parskip 0in
Using CCA, we projected shared embeddings from the speech and language models onto these three spaces. The resulting CC projections were used to create neural encoding models, with contextual CCs demonstrating brain prediction performance similar to the shared components between GPT-2 and Wav2Vec2.0 (Fig. 4b). This performance was significantly higher than that of acoustic-phonetic and static semantic CCs (Fig. 4b). Thus, contextual information, such as structural context, primarily accounts for the shared information between the speech and language models and the brain.

\section{Conclusion}
In this study, we explore shared representations in SSL models trained for speech and language tasks in relation to human speech perception neural activities. We find shared information is primarily contextual, followed by static semantics and acoustic-phonetic. This suggests transformer-based models can learn linguistically relevant information purely through unsupervised learning with acoustic speech signals. These findings highlight convergence in speech contextual representations in SSL models and the brain network involved in speech perception, providing insights into both SSL models and the neural basis of speech and language processing.

\vfill\pagebreak

\label{sec:refs}

\bibliographystyle{IEEEbib}
\bibliography{strings,refs}

\end{document}